\documentclass[osajnl,twocolumn,showpacs,superscriptaddress,10pt]{revtex4-1} 
\usepackage{amsmath,amssymb,graphicx}
\begin{document}

\title{Optical-frequency measurements with a Kerr-microcomb and photonic-chip supercontinuum}

\author{Erin~S.~Lamb}\email{Corresponding author: ecs223@cornell.edu}
\affiliation{National Institute of Standards and Technology (NIST), Boulder, CO, 80305, USA}

\author{David~R.~Carlson}
\affiliation{National Institute of Standards and Technology (NIST), Boulder, CO, 80305, USA}

\author{Daniel~D.~Hickstein}
\affiliation{National Institute of Standards and Technology (NIST), Boulder, CO, 80305, USA}

\author{Jordan~R.~Stone}
\affiliation{National Institute of Standards and Technology (NIST), Boulder, CO, 80305, USA}
\affiliation{Department of Physics, University of Colorado, Boulder, Colorado, 80309, USA}

\author{Scott~A.~Diddams}
\affiliation{National Institute of Standards and Technology (NIST), Boulder, CO, 80305, USA}
\affiliation{Department of Physics, University of Colorado, Boulder, Colorado, 80309, USA}

\author{Scott~B.~Papp}
\affiliation{National Institute of Standards and Technology (NIST), Boulder, CO, 80305, USA}
\affiliation{Department of Physics, University of Colorado, Boulder, Colorado, 80309, USA}

\begin{abstract}
Dissipative solitons formed in Kerr microresonators may enable chip-scale frequency combs for precision optical metrology. Here we explore the creation of an octave-spanning, 15-GHz repetition-rate microcomb suitable for both $f$--$2f$ self-referencing and optical-frequency comparisons across the near infrared.  This is achieved through a simple and reliable approach to deterministically generate, and subsequently frequency stabilize, soliton pulse trains in a silica-disk resonator. Efficient silicon-nitride waveguides provide a supercontinuum spanning 700 to 2100~nm, enabling both offset-frequency stabilization and optical-frequency measurements with $>$100 nW per mode. We demonstrate the stabilized comb by performing a microcomb-mediated comparison of two ultrastable optical-reference cavities.
\end{abstract}

\maketitle 

\section{Introduction}

Microresonator frequency combs hold the promise of taking frequency comb technology outside of the laboratory by providing lower cost, smaller size, microwave-rate, and lower power consumption frequency comb generators than modelocked lasers. Microcombs based on Kerr resonators generate new comb lines through the process of parametric four-wave mixing of a low power continuous-wave (CW) pump laser \cite{Kippenberg:11,DelHaye:07}, and typically operate with repetition rates of 10-to-1000 GHz. Considerable advances have been made in understanding how to generate stable combs utilizing these devices. Importantly, microresonators exhibiting anomalous dispersion have been shown to support single dissipative solitons in numerous material platforms \cite{Herr:14,Yi:15,Brasch:16}, leading to an output pulse train of stable ultrafast pulses. Single solitons are supported in a regime that is normally thermally unstable because the pump laser must be kept at a lower frequency (red-detuned) than the cavity resonance \cite{Carmon:04}. This poses a challenge generating and maintaining solitons, but numerous techniques have been developed to mitigate these difficulties. These techniques include scanning of the pump laser over the resonance before significant heating takes place \cite{Herr:14}, combining a fast scan with an abrupt change in pump power to shift the resonance \cite{Brasch:16,Yi:15}, backwards-tuning of the pump frequency from a multi-soliton state to deterministically delete solitons from the cavity \cite{Guo:17}, thermally tuning the resonance through the use of resistive heaters \cite{Joshi:16}, re-shaping soliton crystal waveforms \cite{Cole:17, DelHaye:16}, careful analysis of the thermo-optic chaos to determine appropriate scans over the resonance \cite{Bao:17}, and active locking of the soliton pulse, either through the power level of the soliton \cite{Yi:16} or through the pump--resonator detuning \cite{Stone:17,Stone:17b}.

\begin{figure}[t!]
\centerline{\includegraphics[width=3.5in]{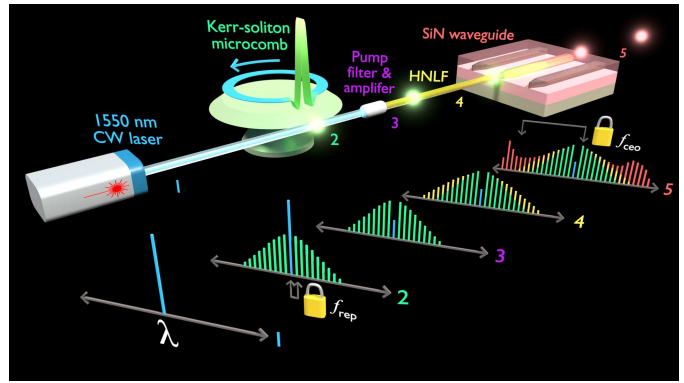}}
\caption{Pictorial representation of frequency comb generation and spectral evolution through the system. A two-stage approach, first using highly-nonlinear fiber (HNLF) and secondly with a silicon-nitride (SiN) waveguide, is used to achieve broadband spectra from the microcomb. Components not altering the comb's spectral shape are omitted from this figure for simplicity. See Figs.~\ref{fig:fig2} and \ref{fig:fig3} for details.}
\label{fig:fig1}
\end{figure}

In order to be useful for precision frequency metrology applications, both the repetition rate of the resonator and offset frequency of the comb must be stabilized. This has traditionally been accomplished by generating an octave-spanning spectrum and frequency doubling the low-frequency end of the spectrum for comparison with its corresponding high-frequency component \cite{Cundiff:03}. Recent experiments have explored self-referenced microcombs, which start with lower pulse energy than their modelocked laser counterparts and thus require additional care to generate full-octave spectra. In prior work from our group, a multi-soliton state was shaped into a single pulse through iterative feedback to a spatial-light modulator (SLM). This pulse was amplified and used to generate a supercontinuum for \mbox{$f$--$2f$} self-referencing in highly nonlinear fiber (HNLF) and was compared to a microwave frequency standard \cite{DelHaye:16}. Jost and coworkers demonstrated a self-referenced magnesium fluoride microresonator by using a single soliton to generate two-thirds of an octave in HNLF and locking external lasers to lines at the edge of the spectrum to perform $2f$--$3f$ self-referencing \cite{Jost:15}. Similarly, $2f$--$3f$ self-referencing with external transfer lasers was performed on two-thirds of an octave generated directly from a silicon nitride resonator utilizing a four-soliton state \cite{Brasch:17}. Recently, another approach to generating an octave-spanning spectrum from a microresonator by using THz-line spacings is being explored \cite{Briles:17,Pfeiffer:17,Spencer:17}. To overcome the difficulty of electronically measuring the terahertz free spectral range (FSR), a novel scheme using a 22~GHz comb to span the terahertz FSR of the octave spanning comb is being developed \cite{Briles:17,Spencer:17}.

In this work, we present results to streamline the self-referencing of both single and double solitons from a 15~GHz silica disk resonator \cite{Lee:12} using traditional \mbox{$f$--$2f$} techniques, as depicted in Figure \ref{fig:fig1}. Since this work relies on directly generating a single or double soliton state with no additional waveform shaping, a key component of our experiment is to reproducibly  generate single solitons and maintain them through active feedback, while providing frequency control of the comb spectrum for repetition rate and carrier--offset frequency stabilization. The active soliton feedback is implemented by monitoring the optical power after the resonator and adjusting the pump laser frequency in order to hold this power level constant, which can keep the soliton pulses stable for more than a day \cite{Stone:17,Stone:17b,Yi:16}.
Then, amplification in an erbium-doped fiber amplifier (EDFA) and pulse compression in normal-dispersion highly nonlinear fiber (HNLF) is utilized to achieve pulses with energies up to 180~pJ and durations of 100 to 120~fs. These pulses are used to generate coherent supercontinuua in silicon-nitride waveguides, which have recently been demonstrated as suitable platforms for allowing frequency comb self-referencing and precision metrology with low-pulse-energy lasers \cite{Mayer:15,Carlson:17,Carlson:17b}.

This use of silicon-chip-based devices for both the soliton generation and octave-plus broadening shows the potential for  chip-integrated sources of broadband, high-repetition-rate frequency combs at the 15~GHz repetition rate, which is useful for its ability to resolve individual comb lines while remaining detectable with microwave electronics. Additionally, using a two-soliton state from the resonator in conjunction with the silicon nitride waveguide, we show a fully self-referenced comb using standard $f$--$2f$ interferometry and use the comb to compare two cavity-stabilized lasers separated by 87 THz across the near-infrared. To our knowledge, this is the first use of a microresonator frequency comb to compare two ultrastable optical cavities. Our work outlines and demonstrates the nonlinear optics and frequency metrology pathway for a self-referenced, chip-integrated microcomb system, provided that ~100 pJ, <100 fs soliton pulses are derived from the Kerr microresonator and subsequent optical amplifcation. Further chip-integration is possible in the future by implementing the pulse compression stage in silicon nitride waveguides.

\begin{figure}[b!]
\centerline{\includegraphics[width=3.5in]{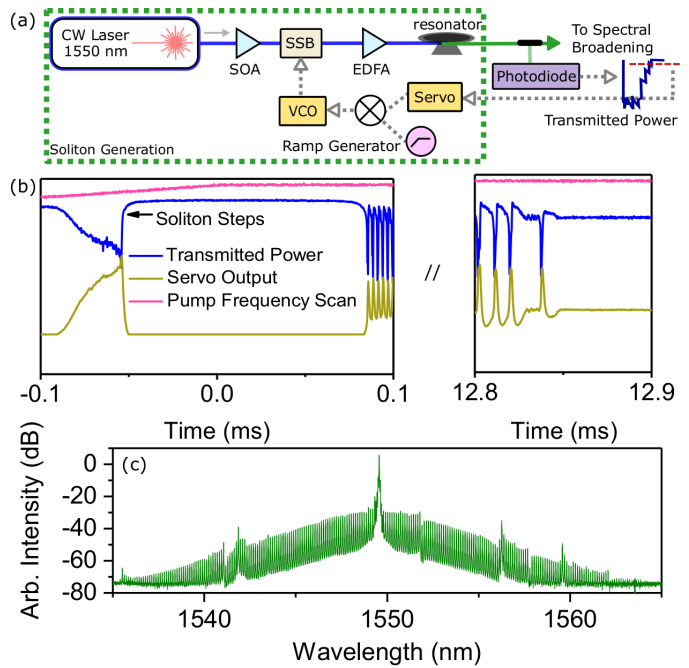}}
\caption{(a) Soliton generation and stabilization using the lock technique described in the text. CW: continuous wave; SOA: semiconductor optical amplifier; SSB: single-side band modulator; EDFA: erbium-doped fiber amplifier; VCO: voltage-controlled oscillator. (b) Experimental frequency scan across the resonance. Multiple scans of the servo over the resonance and the eventual stabilization of a single soliton are shown. (c) Spectrum of a single soliton.}
\label{fig:fig2}
\end{figure}

\section{Soliton generation}

\begin{figure*}[t!]
\centerline{\includegraphics[width=6in]{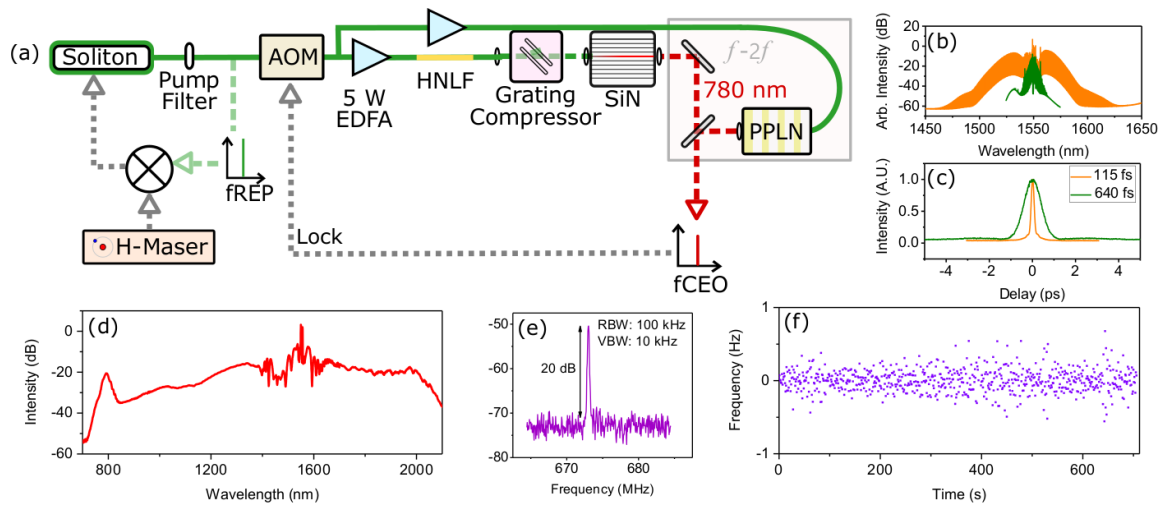}}
\caption{(a) Supercontinuum generation and frequency comb stabilization with a SiN waveguide. fREP: repetition rate frequency; AOM: acousto-optic modulator; HNLF: highly nonlinear fiber; SiN: silicon nitride; PPLN: periodically-poled lithium niobate; fCEO: carrier-envelope offset frequency. (b) Soliton spectrum after pump filter (green) and after broadening in HNLF (orange). (c) Intensity autocorrelation after pump filter (green) and after broadening in HNLF and recompression with a grating pair (orange). (d) Octave-spanning supercontinuum generated when the waveguide is pumped with a single soliton. (e) Locked carrier-envelope offset frequency found by self-referencing the double-soliton state. (f) Counter data for the locked carrier-envelope offset frequency (1~s gate time).}

\label{fig:fig3}
\end{figure*}

We generate single solitons for these experiments using a fast scan across the resonance, followed by stabilization with a servo lock to catch and hold the soliton state \cite{Yi:16,Stone:17}. To implement the fast scan, the pump laser, which is either a wavelength-tunable external-cavity diode laser (ECDL) or a cavity-stabilized fiber laser near 1550~nm, is passed through a single-sideband (SSB) modulator. The SSB modulator contains two parallel Mach-Zehnder interferometers that are driven out of phase and act to suppress the carrier frequency and generate sidebands. Through correct biasing of the interferometers, the carrier and residual sidebands can be suppressed 25 to 30~dB below the first-order sideband, which is used to pump the microresonator after additional amplification through a semiconductor optical amplifier (SOA) and an erbium-doped fiber amplifier (EDFA).

\begin{figure*}
\centerline{\includegraphics[width=7in]{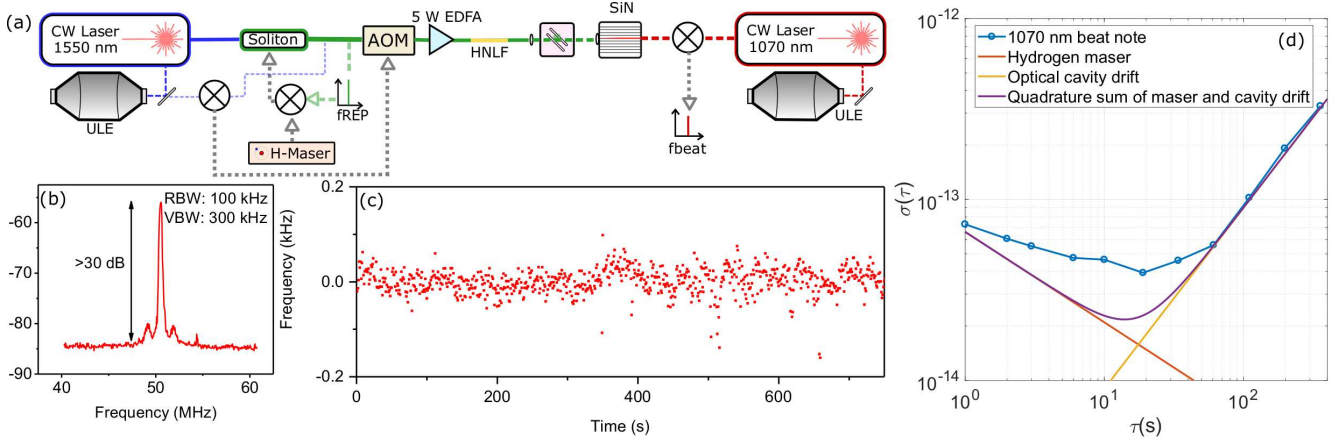}}
\caption{(a) Schematic of optical-frequency comparison measurement between a 1550~nm and a 1070~nm cavity-stabilized laser using the microcomb. The 1550~nm laser is used to pump the microresonator and thus becomes the central tooth of the comb. Frequency fluctuations induced by the soliton lock are canceled using an acousto-optic modulator (AOM) before spectral broadening.  (b) Beat frequency between the 1070~nm laser and the supercontinuum. (c) Counter data of the optical beat shown in (b). (d) The Allan deviation of the data shown in (c), scaled to the 1070-nm carrier, indicates that the stability is limited by the hydrogen maser reference at short time scales and the drift between the optical cavities at longer time scales.}
\label{fig:fig4}
\end{figure*}

High-bandwidth tuning of the SSB frequency can be achieved by programatically steering a voltage-controlled oscillator with a function generator.  In this way the sideband is quickly swept across the cavity resonance and held at a predefined point where a servo lock then captures and holds the soliton by stabilizing the power level corresponding to the desired soliton state (see Fig. \ref{fig:fig2}(a)). A typical scan across the resonance and subsequent lock is shown in Fig. \ref{fig:fig2}(b). Due to the short duration of the soliton steps formed with this 15~GHz device, the initial frequency ramp is rarely precise enough to catch the single soliton on the first scan across the resonance. However, with proper settings of the servo lock, it will repeatedly scan across the resonance, as shown by the oscillations in the servo output in Fig \ref{fig:fig2}(b), until the soliton is stabilized. Using this method, and benefiting from the repeated scanning, the soliton pulse is easily generated and stable over the course of a day. The resulting soliton spectrum obtained directly from the resonator is shown in Figure \ref{fig:fig2}(c).

\section{Supercontinuum generation}

The supercontinuum generation process using the SiN waveguide is depicted in Figure \ref{fig:fig3}(a). The spectrum of the modelocked soliton is modest, so additional pulse compression is used prior to seeding the waveguide. After the resonator, a spatial light modulator (SLM) is used to attenuate the strong pump line, resulting in a smoother spectrum (Fig.~\ref{fig:fig3}(b), green trace) and providing a temporal autocorrelation-width of 640~fs (Fig.~\ref{fig:fig3}(c), green trace). The SLM is also used to compensate for extra dispersion from fiber leads prior to amplification in a 5~W EDFA. The pulse is launched into approximately 5~m of normal dispersion HNLF, which broadens the spectrum through self-phase modulation (SPM) (Fig.~\ref{fig:fig3}(b), orange trace). The pulse is then re-compressed with a grating pair to a duration of 100 to 120~fs (Fig.~\ref{fig:fig2}(c), orange trace). This pulse, which has an energy of around 180~pJ after losses through the grating compressor and fiber connectors, is used to generate supercontinuum light in a low-pressure chemical vapor deposition (LPCVD) silicon-nitride waveguide. The 15~mm waveguide is fully oxide-clad with an approximate thickness of 800~nm and width of 1800~nm. An inverse taper at the input facet reduces coupling losses to less than 2~dB. The use of normal dispersion followed by anomalous dispersion nonlinear media allows a broad supercontinuum to be generated while simultaneously maintaining the coherence and low-noise of the comb \cite{Beha:17}.

When the waveguide is pumped using a single soliton with 100 to 150~pJ of incident pulse energy, a supercontinuum is produced that spans more than an octave (Figure \ref{fig:fig3}(d)). Light from the soliton at 1550~nm is taken prior to the supercontinuum generation and frequency doubled for comparison against the dispersive wave at 775~nm to detect the carrier-envelope offset frequency. Although the offset frequency is easily detectable and can be locked with a single soliton pumping the waveguide, the signal-to-noise ratio on the offset frequency improves when the waveguide is pumped with a double soliton state, presumably due to higher conversion of the CW pump light into soliton pulses in the microresonator. As a practical note, when the pump frequency is decreased by four modes, the offset frequency decreased from 1.1~GHz to the 673~MHz shown in Figure \ref{fig:fig3}(e), giving an offset-frequency tuning capability through selection of the pump frequency. This may enable the offset frequency to be brought into a more easily detectable range.

To achieve full stabilization of the comb, the repetition rate is photodetected and divided by 16 to a frequency of 914~MHz. This frequency is compared to a hydrogen-maser reference and stabilized by electronically adjusting the set-point of the lock on the taper transmission trace used to stabilize the soliton. The offset frequency is detected (as described above) and stabilized through an acousto-optic modulator (AOM) that shifts the microcomb frequencies after the resonator. Using these two actuators, the microcomb can be completely stabilized for over 700~s, as shown in the counter data of the locked offset frequency in Figure \ref{fig:fig3}(f). At longer times, slower thermal drifts of the system prevented maintenance of the lock without adjusting the set-points; this could likely be improved in the future through better thermal stabilization of the system.

\section{Optical cavity comparison}
The microcomb's utility for precision metrology is demonstrated by measuring the relative drift between two ultrastable optical cavities. To implement this measurement (Fig.~\ref{fig:fig4}(a)), a 1550-nm cavity-stabilized laser is passed through the SSB modulator so that the first-order sideband of this laser is used as the pump laser for the microcomb. However, the process of creating the sideband and locking its frequency to hold the soliton induces noise on the pump. The original stability of the cavity-stabilized laser is recovered after soliton generation by detecting a beat between the original light and the microcomb pump line that has been shifted by both the SSB and AOM. This beat is then actively stabilized through feedback to the AOM.  The FSR of the comb is stabilized as before, using a hydrogen-maser reference and actuating the set point of the soliton lock. Then, the double-soliton state is used to generate a supercontinuum in the SiN waveguide, and the resulting supercontinuum line at 1070~nm is beat against a 1070~nm cavity stabilized laser. The beat note, mixed down to 50~MHz, is shown in Fig.~\ref{fig:fig4}(b), along with the corresponding counter data (Fig. \ref{fig:fig4}(c)). These data are converted to an Allan deviation as shown in Fig. \ref{fig:fig4}(d). At short averaging times, the frequency stability is determined by the hydrogen maser reference, and at longer averaging times, the relative linear drift (180~mHz/s) between the two optical cavities dominates. The Allan deviation is larger than the predicted value at around 10~s, which is likely due to uncompensated noise in the several phase locks of the soliton microcomb and possibly in uncompensated path-length noise in our lab experiments that for convenience were here carried over several optics benches.

\section{Noise reduction}

\begin{figure}[htb]
\centerline{\includegraphics[width=\linewidth]{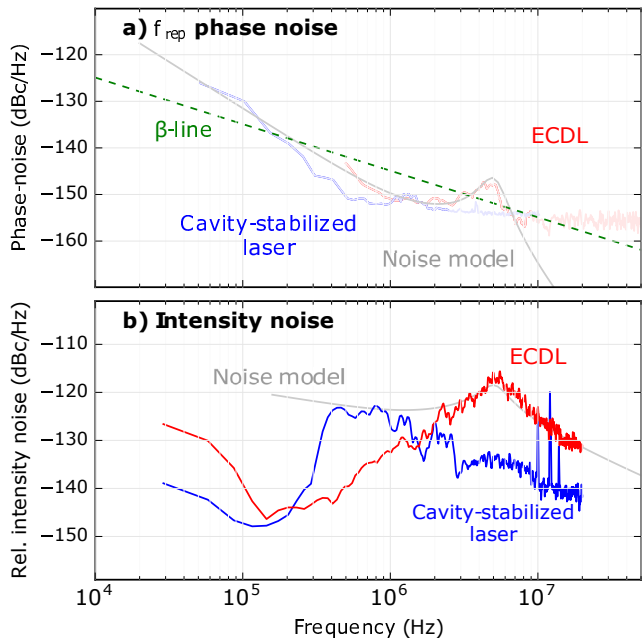}}
\caption{(a) The phase noise of the repetition rate ($f_\mathrm{rep}$) of the broadened microresonator comb, as measured through heterodyne beats at 1070 nm (cavity-stabilized pump laser) and 1111 nm (ECDL pump laser). The ECDL pump produces phase noise above the $\beta$-separation line \cite{DiDomenico:10} at MHz-level frequencies, which degrades the stability of the comb. In contrast, the cavity-stabilized pump laser only exceeds the $\beta$ line below 200~kHz, resulting in lower overall noise. (b) The cavity-stabilized laser also produces much lower intensity noise than the ECDL. The gray line in (a) and (b) is the prediction for the comb noise assuming a pump laser with white frequency noise, as given by Eq. 5 of Ref.~\citenum{Stone:17b}.}
\label{fig:fig5}
\end{figure}

In the process of fully-stabilizing the microresonator, it became
apparent that a high-stability pump laser is critical to producing a low-noise comb. While the full-stabilization described above was completed using a cavity-stabilized laser, early efforts to self-reference the microresonator comb were attempted with an ECDL, which exhibits significantly higher frequency noise. Because a change in the pump laser frequency changes the detuning of the pump laser to the cavity, frequency noise in the pump laser causes intensity noise in the microresonator. As would be expected from this relationship, the noisier ECDL produced much higher levels of relative intensity noise (RIN) at the output of the microresonator compared to the cavity stabilized laser (Fig.~\ref{fig:fig5}(b)). Additionally, since the power in the cavity can affect the repetition rate of the comb via thermal and nonlinear effects \cite{Stone:17b}, RIN in the microresonator causes noise in the frequency of the comb teeth. While the cavity stabilized laser generates a comb with relatively low phase-noise, the ECDL generates a comb with significant phase-noise in the 5-MHz region (Fig.~\ref{fig:fig5}(a)). This phase-noise is above the “$\beta$-line” \cite{DiDomenico:10}, and therefore will be expected to have a strong impact on the stability of the comb and the width of the comb-teeth. We can quantitatively estimate the comb noise using Eq.~5 of Ref.~\citenum{Stone:17b}, which describes the effect that white frequency noise on the pump laser will have on the generated comb. By assuming a detuning of 5~MHz, a coupling of 0.2, a cavity linewidth of 1.9~MHz, and a pump-laser noise level of 80 $\mathrm{Hz/\sqrt{Hz}}$, we can achieve good agreement between the model (gray lines in Fig.~\ref{fig:fig5}) and the experimentally observed phase noise and RIN. The model shows that the phase-noise of the resonator is sensitive to noise at Fourier frequencies at and below the laser--cavity detuning. For future optical frequency metrology experiments with microwave-rate microcomb sueprcontinuua, our work highlights that a low relative frequency noise between the pump laser and the Kerr resonator is critical in achieving a narrow-linewidth and high SNR $f_{CEO}$ signal.

\section{Conclusions}

In conclusion, we have demonstrated reliable soliton generation from a 15~GHz silica disk resonator using a fast scan across the resonance and a servo lock to catch and stabilize the soliton. These solitons are suitable for coherent supercontinuum generation in chip-scale SiN waveguides, enabling the comb to be fully self-referenced using $f$-$2f$ interferometry. The utility of thihs comb for precision metrology is demonstrated by performing, to our knowledge, the first optical frequency comparison using a microcomb.

\section*{Funding Information}

This research is supported by the Air Force Office of Scientific Research (AFOSR) under award number FA9550-16-1-0016, the Defense Advanced Research Projects Agency (DARPA) ACES program, the National Aeronautics and Space Administration (NASA), the National Institute of Standards and Technology (NIST),
and the National Research Council (NRC). This work is a contribution of the U.S. government and is not subject to copyright.

\section*{Acknowledgments}

The authors thank W. Zhang for assistance with the cavity stabilized lasers and T. Briles, D. Spencer, F. Quinlan, and H. Leopardi for useful discussions, comments, and help with the 1070 nm cavity reference laser. We also thank K. Vahala and M. G. Suh for providing the resonator used in these experiments.

\bibliographystyle{osajnl}
\bibliography{MicrocombSoliton}

\end{document}